\documentclass[pmlr]{jmlr}
\usepackage{enumitem}
\usepackage{comment}
\RequirePackage{graphicx}
 \usepackage{booktabs}
\usepackage{longtable}
\usepackage{multirow}
\usepackage{hyperref}

\usepackage{pifont}
\newcommand{\cmark}{\ding{51}}%
\newcommand{\xmark}{\ding{55}}%

\makeatletter
\def\set@curr@file#1{\def\@curr@file{#1}} 
\makeatother
\usepackage[load-configurations=version-1]{siunitx} 

\theorembodyfont{\upshape}
\theoremheaderfont{\scshape}
\theorempostheader{:}
\theoremsep{\newline}

\jmlrvolume{298}
\jmlryear{2025}
\jmlrworkshop{Machine Learning for Healthcare}

\title[INSIGHT: Explainable Weakly-Supervised Medical Image Analysis]{INSIGHT: Explainable Weakly-Supervised Medical Image Analysis}

\author{%
\Name{Wenbo Zhang} \Email{wzhang66@ur.rochester.edu}\\
\addr University of Rochester, Rochester, NY, USA
\AND
\Name{Junyu Chen} \Email{jchen175@ur.rochester.edu}\\
\addr University of Rochester, Rochester, NY, USA
\AND
\Name{Christopher Kanan} \Email{ckanan@cs.rochester.edu }\\
\addr University of Rochester, Rochester, NY, USA
}

\begin{document}

\maketitle

\begin{abstract}
  Due to their large sizes, volumetric scans and whole-slide pathology images (WSIs) are often processed by extracting embeddings from local regions and then an aggregator makes predictions from this set. However, current methods require post-hoc visualization techniques (e.g., Grad-CAM)  and often fail to localize small yet clinically crucial details. To address these limitations, we introduce INSIGHT, a novel weakly-supervised aggregator that integrates heatmap generation as an inductive bias. Starting from pre-trained feature maps, INSIGHT employs a detection module with small convolutional kernels to capture fine details and a context module with a broader receptive field to suppress local false positives. The resulting internal heatmap highlights diagnostically relevant regions. On CT and WSI benchmarks, INSIGHT achieves state-of-the-art classification results and high weakly-labeled semantic segmentation performance. Project website and code are available at: 
\href{https://zhangdylan83.github.io/ewsmia/}{\texttt{https://zhangdylan83.github.io/ewsmia/}}

\end{abstract}

\section{Introduction}

Advances in medical imaging technology have enabled clinicians to extract critical insights from increasingly large and complex datasets. However, the size of medical images presents computational and analytical challenges: whole-slide images (WSIs) of histopathology slides can contain billions of pixels~\citep{campanella2019clinical}, and volumetric scans, such as CT or MRI, are composed of hundreds of slices. Processing such data end-to-end with deep neural networks is computationally infeasible. Instead, pipelines rely on aggregators, which synthesize local embeddings extracted from tiles (WSIs) or slices (volumes) into global predictions~\citep{casson2024joint,chang2022deep,mahmood2021detecting}. While this divide-and-conquer strategy is efficient, current methods often discard spatial information during feature aggregation and depend on post-hoc visualization tools, such as Grad-CAM~\citep{Selvaraju_2019}, to generate interpretable heatmaps. These visualizations are prone to missing clinically significant features and introduce additional complexity. 

\begin{figure*}[t]
	\centering
	\includegraphics[width=0.99\textwidth]{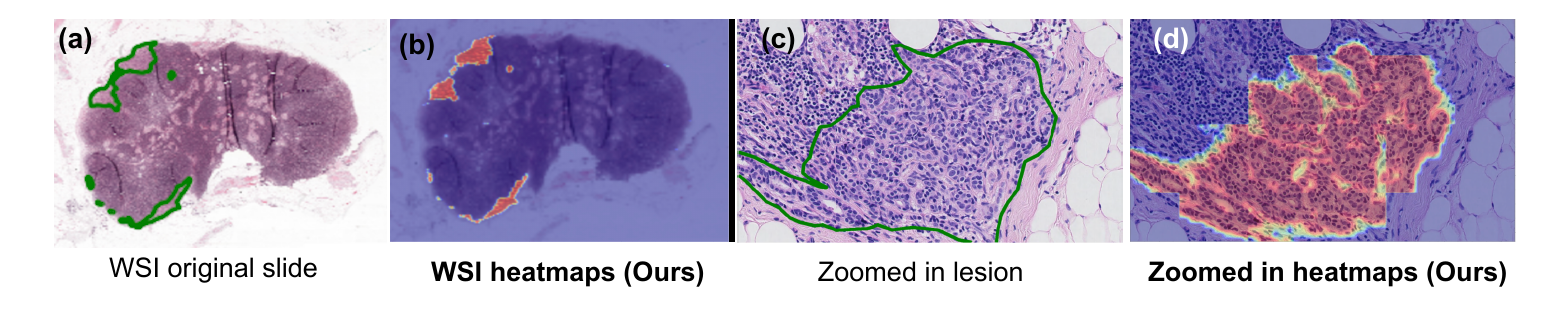}
    \vspace{-5mm}
	\caption{Qualitative visualization of interpretable heatmaps generated by INSIGHT.
(a) Original whole-slide image (WSI) with annotated tumor regions outlined in green. (b) WSI-level heatmap produced by our method, highlighting predicted tumor regions. (c) Zoomed-in view of a representative lesion with expert annotation. (d) Corresponding zoomed-in heatmap showing precise localization of the lesion by INSIGHT. All heatmaps are generated using \textbf{only} image-level categorical labels.} 
    \vspace{-8mm}
	\label{fig:teaser}
\end{figure*}

To address these limitations, we propose \textbf{INSIGHT} (\textbf{I}ntegrated \textbf{N}etwork for \textbf{S}egmentation and \textbf{I}nterpretation with \textbf{G}eneralized \textbf{H}eatmap \textbf{T}ransmission), a novel weakly-supervised aggregator that embeds interpretability directly into its architecture. INSIGHT processes pre-trained patch- or slice-level embeddings through two complementary modules: a \textit{Detection Module}, which captures fine-grained diagnostic features, and a \textit{Context Module}, which suppresses irrelevant activations by incorporating broader spatial information. These outputs are fused to create interpretable \emph{internal} heatmaps that naturally align with diagnostic regions. INSIGHT produces internal segmentation-quality heatmaps \textbf{without} requiring pixel- or voxel-level annotations as shown in Fig.~\ref{fig:teaser}.

Weak supervision is essential in medical imaging, where the creation of detailed annotations is expensive and labor-intensive~\citep{campanella2019clinical}. Fully supervised methods like U-Net~\citep{ronneberger2015unetconvolutionalnetworksbiomedical} and DeepLab~\citep{chen2017deeplabsemanticimagesegmentation} rely on densely labeled data, which limits their scalability. In contrast, multiple instance learning (MIL)~\citep{carbonneau2018multiple,campanella2019clinical} enables models to aggregate features from local regions using only image- or slide-level labels, avoiding the need for pixel-level annotations. This approach allows models to train on significantly larger datasets, as physicians typically diagnose patients without annotating individual image regions. For instance, the FDA-cleared Paige Prostate system demonstrates the clinical viability of weak supervision by achieving high diagnostic accuracy for prostate cancer detection~\citep{zhu2024harnessing,campanella2019clinical}. Similarly, MIL-based methods have effectively aggregated patch- or slice-level features to make predictions on WSIs and volumetric scans~\citep{casson2024joint,chang2022deep}.

However, current weakly supervised methods face critical limitations. They require post-hoc visualization techniques, such as Grad-CAM~\citep{Selvaraju_2019} or class activation maps (CAM)~\citep{zhou2015learningdeepfeaturesdiscriminative}, to produce interpretable saliency maps, adding complexity and often failing to localize small yet clinically essential details. These saliency maps can also be biased by dataset characteristics, such as organ co-occurrence, and often struggle to balance classification accuracy with spatial localization~\citep{cohen2020chesterwebdeliveredlocally}. This leaves considerable room for improvement in designing aggregators that can better capture spatial dependencies and produce clinically meaningful outputs. INSIGHT addresses these challenges by directly embedding interpretability into its architecture, eliminating the reliance on post-hoc methods and bridging the gap between computational efficiency and clinical relevance.

\paragraph{This paper makes the following contributions:}
\begin{itemize}[noitemsep,nolistsep]
    \item We introduce INSIGHT, a novel weakly-supervised aggregator that directly integrates explainability into its architecture, producing high-quality heatmaps as part of its decision-making process.
    \item We demonstrate that INSIGHT achieves state-of-the-art performance on three challenging medical imaging datasets, excelling in both classification and segmentation tasks across CT and WSI modalities using only image-level categorical labels.
    \item Through extensive qualitative and quantitative analyses, we show that INSIGHT’s heatmaps align closely with diagnostic regions, requiring only image-level labels for supervision.
\end{itemize}
\vspace{-2mm}

\subsection*{Generalizable Insights about Machine Learning in the Context of Healthcare}
INSIGHT demonstrates that fine-grained interpretability and strong predictive performance are not mutually exclusive in weakly supervised settings. By maintaining spatial resolution throughout the feature processing pipeline and embedding interpretability directly into the architecture, our method provides a general design principle for scalable and explainable medical AI systems. This architectural approach—preserving locality while incorporating contextual suppression—can be extended beyond pathology and CT to other modalities such as MRI or ultrasound. More broadly, INSIGHT challenges the common reliance on post-hoc attention visualization, offering a viable alternative that integrates decision explanation into the prediction mechanism itself.

For the clinical radiologist or pathologist, INSIGHT produces built-in heatmaps that align with diagnostic regions without requiring pixel-level annotations. These interpretable outputs help clinicians verify model predictions and identify overlooked regions, while traditional pixel-level annotation remains both labor-intensive and time-consuming. By reducing reliance on such dense annotations and producing visual explanations, INSIGHT can accelerate image review and enhance diagnostic confidence, especially for subtle 
or ambiguous
cases.

\begin{figure*}[t] 
  \setlength{\hsize}{\textwidth}
  \vspace{-30pt}
  \centering
\includegraphics[width=0.7\linewidth]{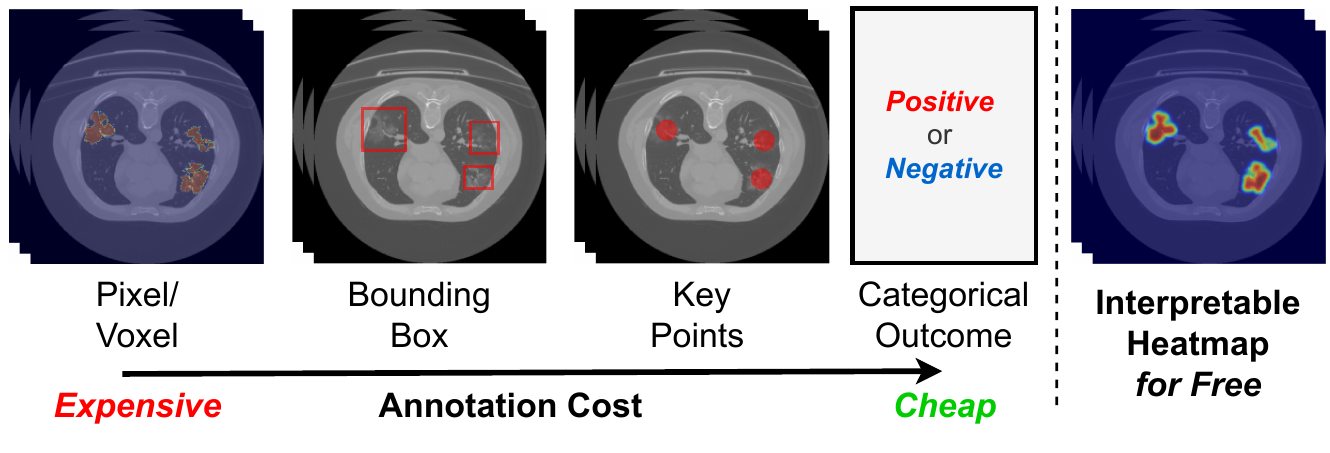}
\vspace{-5mm}
    \caption{Semantic segmentation for medical data often requires expensive annotations that are not done routinely in clinical practice, whereas INSIGHT only needs Categorical labels (left). Unlike other weakly-labeled methods, INSIGHT has per-label heatmaps built into the architecture as a form of inductive bias, enabling it to directly produce heatmaps that closely align with ground-truth diagnostic regions \emph{without} the use of post-hoc methods like Grad-CAM (right).}
    \vspace{-8mm}
    \label{fig:visual-abstract}
\end{figure*}

\section{Related Work}

\subsection{Weakly Supervised Medical Image Learning}
Fully supervised methods such as U-Net~\citep{ronneberger2015unetconvolutionalnetworksbiomedical} and DeepLab~\citep{chen2017deeplabsemanticimagesegmentation} have achieved high segmentation accuracy in medical imaging. However, their reliance on dense, pixel- or voxel-level annotations limits scalability in clinical workflows~\citep{campanella2019clinical}. Weakly supervised learning mitigates this issue by enabling models to train on coarser annotations, such as image- or slide-level labels, which are more practical in large-scale clinical datasets.
Methods based on MIL have proven effective for WSIs and volumetric data, where detailed annotations are unavailable. MIL aggregates features from local patches or slices to predict global outcomes, bypassing the need for pixel-level labels. For example, the FDA-cleared Paige Prostate system demonstrated the clinical viability of MIL by achieving high diagnostic accuracy for prostate cancer detection using only slide-level labels~\citep{campanella2019clinical,zhu2024harnessing}. MIL-based methods have also been successfully applied to breast cancer subtyping from WSIs~\citep{casson2024joint} and chemotherapy response prediction from volumetric CT scans~\citep{chang2022deep}.

While alternative weak labeling strategies, such as scribbles~\citep{LIU2022108341}, bounding boxes~\citep{LU2023109861}, or point-level markers~\citep{laradji2020weakly}, reduce annotation demands, they still require explicit region annotations, which are rarely performed by physicians in clinical practice. INSIGHT builds on the MIL paradigm by utilizing only image-level labels for both classification and segmentation. Unlike existing methods, INSIGHT incorporates interpretability directly into its architecture, generating fine-grained heatmaps that align with diagnostic regions without requiring post-hoc visualization techniques.

\subsection{Aggregation Techniques in Medical Imaging}
Aggregation techniques are essential for efficiently processing large-scale medical imaging data, such as WSIs and volumetric scans, without requiring dense annotations. In WSIs, MIL-based methods treat each slide as a collection of patches and aggregate features for global predictions. For instance, MIL has been successfully applied to cancer detection using slide-level labels~\citep{campanella2019clinical}, while advancements like STAMP~\citep{nahhas2023wholeslideimagebiomarkerprediction} and tissue-graph approaches~\citep{pati2023weaklysupervisedjointwholeslide} have improved WSI analysis by refining aggregation techniques to detect patterns and segment tissues.
In volumetric imaging, slice-level features are aggregated to form volume-level predictions. For example, multi-resolution systems like~\citet{Saha_2020} use DenseVNet-generated features to improve classification of chest CT scans, while~\citet{YE2022108291} developed an aggregation framework with label correction to enhance outcome predictions.

INSIGHT advances these techniques by preserving spatial resolution across slices or patches until the final pooling stage, enabling the generation of detailed, localized heatmaps. This addresses a key limitation of traditional aggregation methods, which often discard spatial information, reducing their ability to localize diagnostically relevant regions.
Existing methods aim at enhancing contextual understanding in feature aggregation, but they each have inherent limitations. For instance, FPN~\citep{lin2017feature} is designed for hierarchical multi-scale feature fusion but lacks an explicit mechanism to independently optimize local and global features, which is crucial for WSI. On the other hand, Dual Attention Network~\citep{fu2019dual} relies on self-attention mechanisms to model long-range dependencies across spatial and channel dimensions, leading to quadratic complexity and computationally expensive for high-resolution medical images such as WSI. In contrast, unlike FPN which depends on predefined hierarchical feature aggregation, INSIGHT's Context module efficiently refines activations within the same feature scale. It also allows separate optimizations on local and global representations. Meanwhile, compared to Dual Attention Nets, INSIGHT avoids the quadratic complexity of self-attention with larger convolution kernel size, making it significantly more efficient for processing whole-slide images.

\subsection{Explainable AI Methods in Medical Imaging}
Explainable AI is crucial for ensuring model predictions in medical imaging are interpretable and clinically meaningful. Post-hoc methods such as class activation maps (CAM)~\citep{zhou2015learningdeepfeaturesdiscriminative} and Grad-CAM~\citep{Selvaraju_2019} are widely used to generate saliency maps that highlight regions associated with model predictions. For example,~\citet{XIE2023102771} developed high-resolution CAMs for thoracic CT scans, and~\citet{Viniavskyi2020} refined pseudo-labels for chest X-ray segmentation using CAMs.
Attention mechanisms have further advanced interpretability. Self-attention, as introduced in~\citep{vaswani2023attentionneed}, has been combined with Grad-CAM to localize disease-specific regions~\citep{Zhang_2022}, while dual-attention mechanisms, such as DA-CMIL \citep{chikontwe2021dual}, produce interpretable maps for detecting conditions like COVID-19 and bacterial pneumonia.

\begin{figure}[t]
    \centering
    \includegraphics[width=\linewidth]{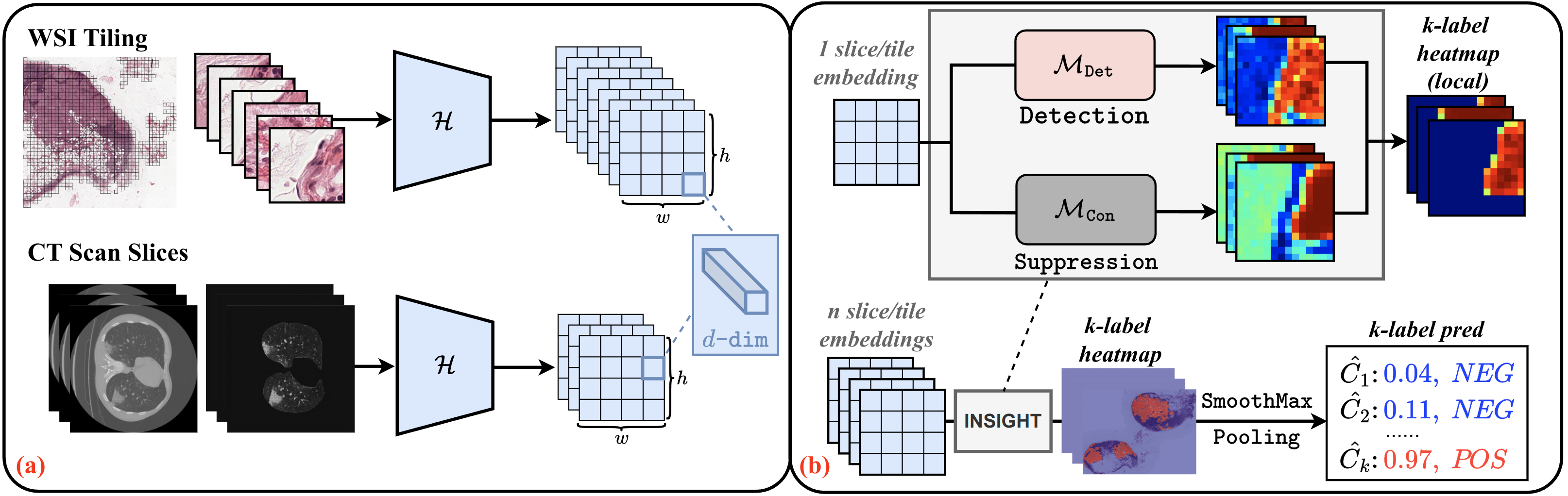}
    \vspace{-5mm}
    \caption{Overview of the INSIGHT framework. \textbf{(a)} Input images (WSIs or CT volumes) are preprocessed and transformed into spatial embeddings using a pretrained encoder. \textbf{(b)} Each slice or patch embedding is processed by INSIGHT, which consists of a detection module to capture fine-grained signals and a context suppression module to reduce false positives. This produces patch-level heatmaps for each category, which are then aggregated to generate whole-slide or volume-level heatmaps. Final predictions for each category are obtained via SmoothMax pooling over the heatmaps. Throughout this process, spatial resolution is preserved, encouraging the model to produce reliable and interpretable heatmaps.}
    \vspace{-5mm}
    \label{fig:model}
\end{figure}

While these methods generate useful insights, they rely on post-hoc interpretability, which can misalign with the model's decision-making process. Few existing methods directly integrate interpretability into their model architectures. One example is Shallow-ProtoPNet~\citep{singh2025shallowest}, a prototype-based network designed for fully transparent inference on small-scale datasets such as chest X-rays. In contrast, INSIGHT is built to scale to large, high-resolution inputs like CT volumes and WSIs, addressing the computational and spatial challenges that arise in real-world clinical imaging. INSIGHT eliminates this reliance by embedding interpretability directly into its architecture, producing calibrated, built-in heatmaps as part of its predictions, as shown in Fig.~\ref{fig:model}.  ensures alignment between the model's outputs and diagnostic reasoning.

\subsection{Pre-training in Medical Image Analysis}
Pre-trained models have significantly advanced medical imaging by providing robust feature extractors. ResNet-based architectures, such as those used in~\citep{courtiol2020classificationdiseaselocalizationhistopathology,lin2022identifying}, effectively capture local and intermediate features in weakly supervised frameworks. 

On the other hand, Vision Transformers (ViTs)~\citep{dosovitskiy2021imageworth16x16words} leverage self-attention to capture broader spatial relationships. Self-supervised foundational models are typically pre-trained on massive unlabeled image datasets, enabling the learning of meaningful representations without manual annotations. They produce high-quality features that are highly compatible with downstream tasks including those based on aggregation. 

For example, UNI~\citep{chen2024uni} is trained on over $100$ M WSIs using a DINOv2-style~\citep{oquab2024dinov2} self-supervised learning framework and provides pathology-specific representations optimized for fine-grained localization. Virchow 2~\citep{zimmermann2024virchow2scalingselfsupervisedmixed}, a $632$-million-parameter ViT-H model, is pre-trained on $3.1$ M WSIs with domain-specific augmentations tailored for histopathology tasks.
INSIGHT leverages pre-trained ViTs to extract tensor embeddings from WSI patches and radiology slices. By preserving spatial resolution, INSIGHT achieves robust performance across WSI and CT datasets, supporting fine-grained analysis and interpretability.

\section{Methods}

Fig.~\ref{fig:model} illustrates the high-level overview architecture of INSIGHT. INSIGHT processes pre-trained features from each slice in a volume or each patch in a WSI through dual modules—Detection and Context—to generate fine-grained heatmaps. These heatmaps are aggregated using a SmoothMax pooling strategy to produce categorical predictions, optimized using a combination of binary cross-entropy (BCE) and spectral decoupling losses. Below, we describe each component in detail.

\subsection{Feature Extraction}
To preserve spatial resolution and capture both local and global contextual information, INSIGHT uses pre-trained models specific to the input modality, as illustrated in Fig~\ref{fig:model}(a). To demonstrate that INSIGHT is compatible with a variety of foundation models across modalities, we evaluate it with multiple state-of-the-art encoders.
For CT scan volumes, we use a ViT pre-trained with DINOv2~\citep{oquab2024dinov2} on ImageNet. For pathology WSIs, we evaluate INSIGHT with both UNI~\citep{chen2024uni} and Virchow 2~\citep{zimmermann2024virchow2scalingselfsupervisedmixed}. UNI is pre-trained using DINOv2-style self-supervision on over 100 million WSIs, enabling it to capture rich visual features tailored to histopathology. Virchow 2 is a large ViT-H model with 632 million parameters, trained on 3.1 million WSIs using domain-specific augmentations optimized for pathology. In all cases, the extracted feature maps are denoted as $F \in  \mathbb{R}^{c \times h \times w}$, where $c$, $h$, and $w$ are the feature depth, height, and width, respectively.

\subsection{INSIGHT Architecture}

\noindent \textbf{Heatmap Fusion.} The core of INSIGHT lies in its architectural design which introduces an inductive bias that integrates both local and global spatial features. As illustrated in Fig.~\ref{fig:model}(b), each input embedding is processed by two parallel modules: the \textit{Detection module} and the \textit{Context modules}, which are combined to generate a refined heatmap:
\begin{equation}
    \mathcal{H} = \sigma((1-\sigma(\mathcal{H}_\texttt{Con})) \odot \mathcal{H}_\texttt{Det}),
\end{equation}
where $\sigma$ is the sigmoid activation, and $\odot$ denotes element-wise multiplication. The resulting heatmap $\mathcal{H}$ assigns relevance scores to regions, aligning with the target labels.
\textit{Detection module} uses small-kernel convolutions to capture fine-grained details such as textures and edges, producing a heatmap $\mathcal{H}_\texttt{Det}$ that highlights potential regions of interest. To mitigate false positives, the \textit{Context module} uses larger kernels to capture broader spatial context and suppress irrelevant activations. To better align the pre-trained features with the downstream task, we apply a linear projection before passing the embeddings to both modules.

To isolate high-saliency regions within the fused heatmap, we apply dynamic thresholding with Otsu’s algorithm~\citep{otsu1979threshold}:
\begin{equation}
   \mathcal{H^\prime} = \mathcal{H} \cdot \mathbb{I}(\mathcal{H} > T), 
\end{equation}
where $T = \arg \min_t \sigma_w^2(t)$ minimizes intra-class variance and $\mathbb{I}$ is the indicator function. This step enhances localization by filtering out low-activation background regions.

\noindent \textbf{Pooling and Classification. }
To produce a categorical outcome, heatmaps are aggregated across all slices or patches using SmoothMax pooling~\citep{maddison2016concrete}, which is a Boltzmann-weighted operator that emphasizes regions with strong activations:
\begin{equation}
  \hat{y} = \frac{\sum_{i} \mathcal{H^\prime}_i \cdot \exp(\alpha \cdot \mathcal{H^\prime}_i)}{\sum_{i} \exp(\alpha \cdot \mathcal{H^\prime}_i)},
\end{equation}
where $\hat{y}$ is the predicted likelihood of a positive outcome, $\mathcal{H^\prime}_i$ are heatmap activations, and $\alpha$ controls the sharpness of the weighting. For multi-label classification, the heatmap is extended across channels, with each channel representing a separate category, allowing joint optimization over multiple labels.

We empirically tested several other pooling methods, including max pooling and LP pooling to compare their effectiveness. Max pooling considers only the strongest activation, which limits spatial gradient flow during training and leads to suboptimal heatmap quality. LP pooling (\( p \geq 2 \))incorporates more values but often produces outputs outside the 
$[0,1]$ range, requiring post-processing (e.g., output clipping or sparse regularization~\citep{zhou2021learning}), which yielded no empirical gains.  In contrast, SmoothMax pooling provides a principled trade-off between local emphasis and global context through exponential weighting.

\subsection{Training Objective}
\noindent \textbf{Binary Cross-Entropy Loss.}
Since our model is trained using only volume- or slide-level categorical labels, we optimize the predicted probabilities using the binary cross-entropy loss $\mathcal{L}_\text{BCE}$. For multi-label settings, the loss is independently computed for each class.

\noindent \textbf{Spectral Decoupling.} To address gradient starvation and encourage robust decision boundaries, we include spectral decoupling loss~\citep{pohjonen2022spectral}:
\begin{equation}
   \mathcal{L}_{\text{SD}} = \frac{\lambda_{\text{SD}}}{2} \cdot \|z\|_2^2,
\end{equation}
where $z$ are unnormalized logits, and $\lambda_{\text{SD}}$ controls regularization strength. Spectral decoupling is one of the most effective methods for overcoming dataset bias without additional sub-group labels~\citep{shrestha2022investigation,shrestha2022occamnets}.

\noindent \textbf{Total Loss.} The total loss combines these objectives:
\begin{equation}
\mathcal{L} = \mathcal{L}_{\text{BCE}} + \mathcal{L}_{\text{SD}}.
\end{equation}

\subsection{Heatmap Visualization}
Heatmaps are reconstructed to match the original image dimensions for intuitive interpretation. For WSIs, patch-level heatmaps are stitched together based on their spatial coordinates. For CT volumes, slice-level heatmaps are stacked in order to reconstruct the full volume. The resulting heatmaps are then upsampled using bicubic interpolation to ensure smooth and visually consistent representations.
The pseudocode for the heatmap generation and classification process is provided in Appendix~\ref{appendix:algorithm}.

\section{Datasets and Preprocessing}
\label{sec:prepro}
We evaluate INSIGHT on three publicly available datasets representing two imaging modalities: whole slide images (WSI) and CT volumes. 

\subsection{Computed Tomography Dataset} 
\textbf{MosMed}~\citep{morozov2020mosmeddatachestctscans}. The MosMed dataset comprises $1,110$ chest CT studies for COVID-19 detection, split into two subsets: MosMed-A ($1,060$ volumes) and MosMed-B ($50$ volumes). MosMed-A is used for classification tasks, employing five-fold cross-validation with an $80$/$20$ train-test split. MosMed-B, containing voxel-level annotations, is used solely to evaluate segmentation performance by comparing INSIGHT-generated heatmaps $\mathcal{H}$ against ground truth using Dice score.
CT volumes are normalized to the Hounsfield Unit (HU) range $[-1000, 400]$, scaled to $[0, 1]$, and resized to $518 \times 518 \times 32$. To ensure compatibility with pre-trained ViTs, single-channel data is replicated across three channels. Lung parenchyma is isolated using Lungmask~\citep{hofmanninger2020automatic}.

\subsection{Whole Slide Image Datasets}
\label{sec:wsi-data}
\textbf{CAMELYON16}~\citep{litjens2018camelyon}. This dataset contains $399$ WSIs for detecting metastatic breast cancer in lymph nodes. Following UNI~\citep{chen2024uni}, we use the $20\times$ magnification slides in our experiments, omitting one unusable slide from the original training set. We split the remaining $269$ training slides into $243$ for training and $26$ for validation. The model checkpoint with the highest validation Dice is evaluated on the official test set of $129$ slides.

\noindent \textbf{BRACS}~\citep{brancati2021bracsdatasetbreastcarcinoma}. This dataset contains $547$ WSIs annotated with $4539$ bounding boxes for breast carcinoma subtype classification: \textit{ADH} (Atypical Ductal Hyperplasia – early premalignant change), \textit{FEA} (Flat Epithelial Atypia – precursor lesion), \textit{DCIS} (Ductal Carcinoma in Situ – non-invasive cancer), and \textit{Invasive} (Invasive Carcinoma – malignant stromal invasion). As BRACS provides only the most severe label for each slide, we extracted fine-grained labels using QuPath~\citep{bankhead2017qupath}. Following BRACS’s official splits and excluding slides with incomplete labeling or no valid regions, the final dataset comprises $347$, $51$, and $73$ slides for training, validation, and testing. We downsample all WSIs from $40\times$ to $20\times$ magnification.

We use the CLAM toolbox~\citep{lu2021data} for patch extraction for both CAMELYON16 and BRACS.

\section{Experiments} 
This section presents the experimental setup and results. We provide additional implementation details, including training configurations and model architecture parameters, in Appendix~\ref{appendix:training}.

\subsection{Compared Baselines}
\label{sec:baseline}
We compare INSIGHT to state-of-the-art methods across the MosMed, CAMELYON16, and BRACS datasets, evaluating both classification and segmentation performance under a weakly supervised setting.

\noindent \textbf{CT Volumes.} 
For the MosMed dataset, we benchmark against methods specifically designed for COVID-19 CT analysis. 

\begin{itemize}
    \item \textbf{Progressively Resized 3D-CNN}~\citep{hasan2021covid19identificationvolumetricchest}, a CNN that progressively resizes input volumes to extract multi-scale information.
    \item \textbf{3D U-Net}~\citep{bressem20213dunetsegmentationcovid19}, a fully supervised segmentation model trained with voxel-level annotations that provides a direct comparison compared to weakly supervised methods.
    \item \textbf{3D GAN}~\citep{shabani2022self}, a weakly supervised generative model that performs volume-level segmentation without requiring dense voxel-level annotations.
\end{itemize}

\noindent \textbf{WSI Datasets.}
To ensure a fair comparison, all methods (including ours) use \textbf{the same foundational model} encoder for feature extraction:
\begin{itemize}
    \item \textbf{ABMIL}~\citep{ilse2018attentionbaseddeepmultipleinstance}: An attention-based MIL approach that identifies the most informative regions within a slide, commonly used in weakly supervised WSI analysis.

    \item \textbf{CLAM-SB/MB}~\citep{lu2021data}: A MIL method combining attention mechanisms with clustering to improve localization. CLAM-SB uses a single branch for non-overlapping classes, while -MB employs multiple branches to target specific classes.

    \item \textbf{TransMIL}~\citep{shao2021transmiltransformerbasedcorrelated}: A transformer-based MIL model that captures long-range dependencies within WSIs, enhancing contextual analysis for slide-level classification.

    \item \textbf{WiKG}~\citep{li2024dynamic}: A dynamic graph representation learning framework that models knowledge graphs, implementing a knowledge-aware attention mechanism to capture contextual interactions.
\end{itemize}

\subsection{Quantitative Results}
INSIGHT consistently outperformed baseline methods in classification and segmentation tasks across all datasets, demonstrating adaptability to diverse medical imaging modalities. Performance metrics are reported in Table~\ref{table:mosmed_performance} and Table~\ref{table:combined_performance}. 
On \textbf{MosMed}, INSIGHT achieved a classification AUC of 96.2\%, nearly 5\% higher than the best baseline, underscoring its robustness in volumetric CT analysis for COVID-19 detection.
On \textbf{CAMELYON16}, INSIGHT achieved a Dice score of 74.6\%, outperforming the top competing model by 6.9\%, highlighting its ability to enhance boundary precision in segmentation.
On \textbf{BRACS}, INSIGHT showed strong multi-label classification performance, with AUC improvements of 3.3\% for both ADH and FEA, showcasing its effectiveness in distinguishing subtle tissue variations critical for subtype classification.

INSIGHT’s superior performance stems from three key architectural innovations:

\noindent \textbf{Spatial Embedding.} By retaining spatial resolution during feature extraction, INSIGHT produces $14 \times 14 \times 1024$ feature maps per WSI tile, preserving diagnostically meaningful structure. In contrast, baseline methods reduce feature maps to $1 \times 1 \times 1024$ via pooling, discarding fine-grained information. This enables INSIGHT to detect small but critical patterns, especially in BRACS, where distinct subtypes may co-exist within a single slide.

\begin{table}[t]
\centering
\caption{Performance comparison for classification and segmentation on the MosMed dataset. Methods marked with \textsuperscript{*} were trained on an external dataset but tested on the same MosMed test set for consistency. Baseline details in Sec.~\ref{sec:baseline}.}
\renewcommand{\arraystretch}{1.2} 
\setlength{\tabcolsep}{10pt} 
\small 
\resizebox{0.6\textwidth}{!}{
\begin{tabular}{l|l|c}
\toprule
\textbf{Task} & \textbf{Method} & \textbf{AUC / Dice} (\%) \\
\midrule
\multirow{2}{*}{Classification} 
& PR-3D-CNN (Five-fold CV) & 0.914 ± {\scriptsize 0.049} \\
& \textbf{INSIGHT (Five-fold CV)} & \textbf{0.962 ± {\scriptsize 0.012}} \\
\midrule
\multirow{3}{*}{Segmentation}
& 3D U-Net$^*$ (Voxel-level) & 40.5 ± {\scriptsize 21.3} \\
& 3D GAN (Volume-level) & 41.2 ± {\scriptsize 14.7} \\
& \textbf{INSIGHT (Volume-level)} & \textbf{42.7 ± {\scriptsize 15.3}} \\
\bottomrule
\end{tabular}
}
\label{table:mosmed_performance}
\end{table}

\begin{table}[t]
\caption{Performance comparison on the CAMELYON16 and BRACS official test datasets. ALL aggregators use \textbf{UNI} for embeddings. We report AUC and Dice scores (mean ± std) for CAMELYON16, and AUC scores for different categories in BRACS. Reported Dice scores for all comparison methods were fine-tuned through grid search on the validation set. Baseline details in Sec.~\ref{sec:baseline}. BRACS's subtype definitions are provided in Sec.~\ref{sec:wsi-data}.}\centering
\renewcommand{\arraystretch}{1.2}
\setlength{\tabcolsep}{10pt}          
\small
\resizebox{0.7\textwidth}{!}{
\begin{tabular}{l|cc|cccc}
\toprule
\multirow{2}{*}{\textbf{Aggregator}} 
 & \multicolumn{2}{c|}{\textbf{CAMELYON16}} 
 & \multicolumn{4}{c}{\textbf{BRACS}} \\
\cmidrule(lr){2-3}\cmidrule(lr){4-7}
& \textbf{AUC} & \textbf{Dice} 
& \textbf{ADH} & \textbf{FEA} & \textbf{DCIS} & \textbf{Invasive}  \\
\midrule
ABMIL   & 0.975 & 55.8$\pm$25.0 & 0.656 & 0.744 & 0.804 & 0.995 \\
CLAM-SB & 0.966 & 64.7$\pm$24.1 & 0.611 & \underline{0.757} & \underline{0.833} & \underline{\textbf{0.999}} \\
CLAM-MB & 0.973 & \underline{67.7$\pm$22.6} & \underline{0.701} & 0.687 & 0.828 & 0.998 \\
TransMIL& \underline{0.982} & 12.4$\pm$22.4 & 0.644 & 0.653 & 0.769 & 0.989 \\
WiKG    & 0.967 & 66.1$\pm$24.2 & 0.454 & 0.653 & 0.771 & 0.990 \\
\midrule
\textbf{INSIGHT} 
& \textbf{0.990} & \textbf{74.6$\pm$19.1} 
& \textbf{0.734} & \textbf{0.790} & \textbf{0.837} & \textbf{0.999} \\
\bottomrule
\end{tabular}
}
\label{table:combined_performance}
\end{table}

\noindent \textbf{Adjacent Context Focus.} INSIGHT’s context module captures local dependencies using larger convolutional filters. This focus on adjacent spatial relationships is especially effective for multi-label datasets like BRACS, where lesion subtypes often appear in close proximity.

\noindent \textbf{Broader Foreground Activation Consideration.} Unlike MIL-based models that focus only on top \( k \) patches, INSIGHT’s smooth pooling considers all relevant activations above a threshold. This preserves \textit{finer} boundary details, enhancing segmentation accuracy. For example, on CAMELYON16, this broader inclusion contributed to its 6.9\% Dice improvement.

We further conducted experiments using the \textbf{Virchow 2} foundational model~\citep{zimmermann2024virchow2scalingselfsupervisedmixed}, a more recent state-of-the-art pretraining approach for histopathology images, to assess INSIGHT's performance and robustness against other benchmark methods. As shown in Table~\ref{tab:virchow2_results_table}, INSIGHT again largely outperformed all baselines in segmentation task on the CAMELYON16 dataset. INSIGHT’s performance improvements are further supported by qualitative results (Fig.~\ref{fig:qualitative}) and ablation studies (Sec.~\ref{sec:ablation-studies}), demonstrating the critical role of its architectural innovations in achieving state-of-the-art results.

\begin{table}[t]
    \centering
    \caption{Performance comparison on the CAMELYON16 official test dataset using \textbf{Virchow 2} for embeddings. We report AUC and Dice scores (mean ± std) for CAMELYON16. Reported Dice scores for all comparison methods were fine-tuned through grid search on the validation set. Baseline details in Sec.~\ref{sec:baseline}.}    
    \renewcommand{\arraystretch}{1.2}
    \small
    \resizebox{0.35\textwidth}{!}{
    \begin{tabular}{lcc}
        \toprule
        \textbf{Aggregator} & \textbf{AUC} & \textbf{Dice (\%)} \\
        \midrule
        ABMIL     & 0.963 & 59.8$\pm$25.7 \\
        CLAM-SB   & \underline{0.992} & \underline{66.8$\pm$25.1} \\
        CLAM-MB   & 0.986 & 66.2$\pm$21.9 \\
        TransMIL  & 0.980 & 9.3$\pm$20.1 \\
        WiKG      & \textbf{0.996} & \underline{66.8$\pm21.4$} \\
        \textbf{INSIGHT} & \underline{0.992} & \textbf{78.3$\pm$15.4} \\
        \bottomrule
    \end{tabular}
    }
    \label{tab:virchow2_results_table}
\end{table}

\begin{table}[t]
\centering
\caption{Performance comparison (Dice, mean ± std) on different lesion sizes. Paired one-tailed permutation hypothesis tests (10{,}000 iterations) compare each baseline to INSIGHT. Bolded indicates $0.01\leq p < 0.05$, and $^{**}$ indicates $p < 0.01$.}
\label{tab:baseline_dice_p}
\resizebox{0.8\textwidth}{!}{%
\begin{tabular}{llcccccc}
\toprule
& & \multicolumn{2}{c}{\textbf{Small}} 
  & \multicolumn{2}{c}{\textbf{Moderate}} 
  & \multicolumn{2}{c}{\textbf{Large}} \\
\cmidrule(lr){3-4} \cmidrule(lr){5-6} \cmidrule(lr){7-8}
\textbf{Model} & \textbf{Aggregator} 
& \textbf{Dice (\%)} & \textbf{\emph{p}-value}
& \textbf{Dice (\%)} & \textbf{\emph{p}-value}
& \textbf{Dice (\%)} & \textbf{\emph{p}-value} \\
\midrule
\multirow{6}{*}{UNI}
& \textbf{INSIGHT}  & $\mathbf{42.8 \pm 38.6}$ & - & $\mathbf{74.6 \pm 27.3}$ & - & $\mathbf{82.1 \pm 19.8}$ & - \\
\cmidrule(lr){2-8}
& CLAM-MB   & $20.7 \pm 32.2$ & $^{**}$ & $62.2 \pm 27.8$ & $^{**}$ & $\mathbf{81.4 \pm 19.6}$ & $0.40$ \\
& CLAM-SB   & $20.8 \pm 32.2$ & $^{**}$ & $62.6 \pm 27.4$ & $^{**}$ & $\mathbf{81.2 \pm 19.2}$ & $0.36$ \\
& ABMIL     & $11.0 \pm 25.2$ & $^{**}$ & $51.1 \pm 29.8$ & $^{**}$ & $72.4 \pm 23.0$ & $^{**}$ \\
& TransMIL  & $30.6 \pm 37.3$ & $^{**}$ & $23.4 \pm 32.0$ & $^{**}$ & $34.1 \pm 32.3$ & $^{**}$ \\
& WiKG      & $19.5 \pm 31.9$ & $^{**}$ & $58.7 \pm 29.1$ & $^{**}$ & $\mathbf{80.3 \pm 21.0}$ & $0.29$ \\
\midrule
\multirow{6}{*}{Virchow2}
& \textbf{INSIGHT}  & $\mathbf{55.5 \pm 37.0}$ & - & $\mathbf{78.6 \pm 22.1}$ & - & $\mathbf{81.2 \pm 19.7}$ & - \\
\cmidrule(lr){2-8}
& CLAM-MB   & $24.5 \pm 33.3$ & $^{**}$ & $60.9 \pm 24.7$ & $^{**}$ & $\mathbf{81.4 \pm 17.7}$ & $0.52$ \\
& CLAM-SB   & $27.7 \pm 34.3$ & $^{**}$ & $66.6 \pm 22.8$ & $^{**}$ & $\mathbf{84.1 \pm 16.5}$ & $0.81$ \\
& ABMIL     & $10.8 \pm 24.4$ & $^{**}$ & $50.9 \pm 30.1$ & $^{**}$ & $74.6 \pm 23.6$ & $0.04$ \\
& TransMIL  & $42.4 \pm 37.0$ & $^{**}$ & $35.4 \pm 32.8$ & $^{**}$ & $50.9 \pm 29.9$ & $^{**}$ \\
& WiKG      & $24.1 \pm 33.9$ & $^{**}$ & $62.1 \pm 26.4$ & $^{**}$ & $\mathbf{82.8 \pm 17.8}$ & $0.68$ \\
\bottomrule
\end{tabular}}
\end{table}

\subsubsection{Stratified Analysis}
To assess performance across lesion scales, we extracted connected lesion components and corresponding predicted heatmaps from the entire test set. Lesions were grouped into three size-based bins based on pixel area: \textbf{Small} ($<10^6$ pixels), \textbf{Moderate} ($10^6$–$10^7$ pixels), and \textbf{Large} ($>10^7$ pixels). These thresholds were chosen based on the empirical area distribution in the dataset to 
reflect clinically meaningful differences.  The bin counts are: $1415$ (Small), $135$ (Moderate), and $62$ (Large), showing that small lesions dominate the dataset and represent a particularly challenging and clinically important subgroup.
Table~\ref{tab:baseline_dice_p} reports the Dice scores of INSIGHT and comparison models scores across different lesion sizes using both UNI and Virchow 2 backbones, along with one-tailed permutation test results (10,000 iterations) comparing INSIGHT to each baseline. INSIGHT consistently outperforms all baselines on \textbf{Small} and \textbf{Moderate} lesions ($p < 0.01$), which are especially critical in clinical diagnosis due to their subtle presentation. For \textbf{Large} lesions, INSIGHT performs comparably or better than baselines. These findings highlight INSIGHT’s robustness across lesion scales and its particular advantage in detecting small, subtle pathological regions often missed by conventional MIL-based methods.

\subsection{Qualitative Analysis}
\begin{figure*}[t]
	\centering
	\includegraphics[width=0.9\textwidth]{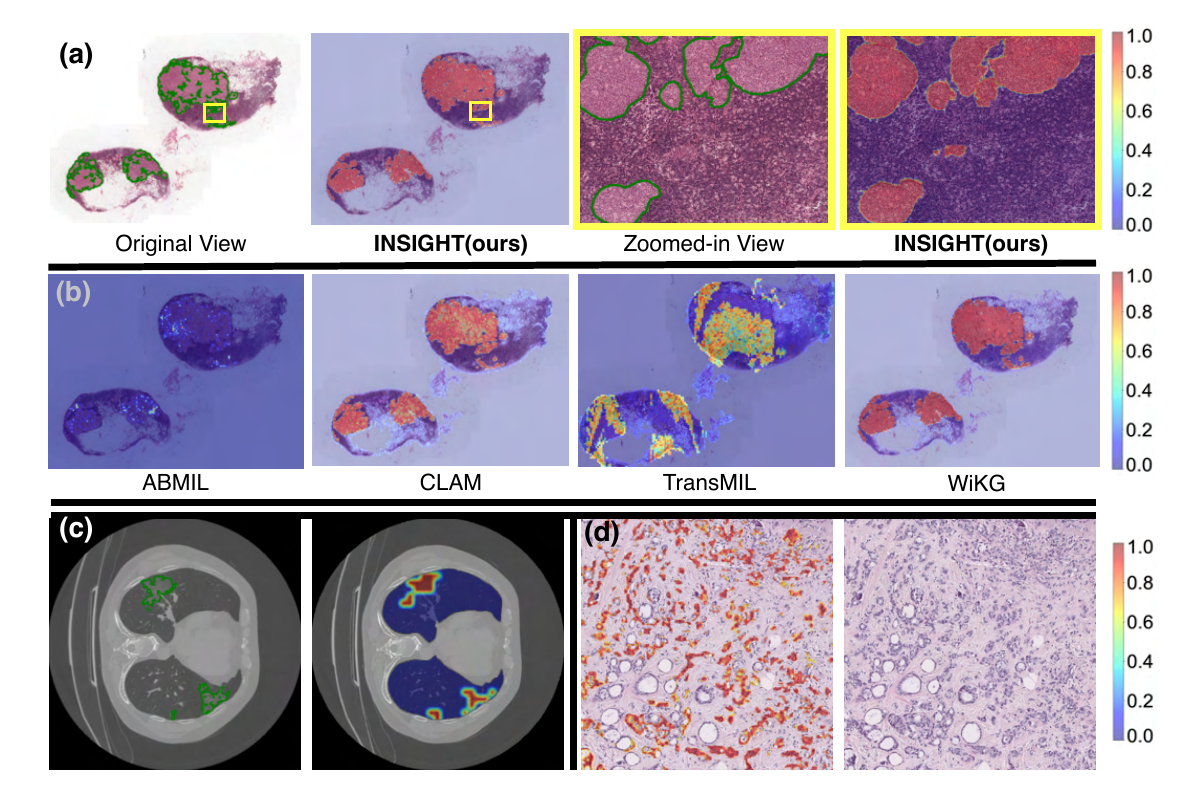}
    \vspace{-5mm}
	\caption{Qualitative comparison of heatmaps generated by INSIGHT (ours) and baseline methods using \textbf{UNI}. For CAMELYON16, we compare lesion detection heatmaps generated by (a) our INSIGHT model and (b) the baseline methods. Additionally, we present (c) heatmaps for MosMed and (d) heatmaps for BRACS. Ground-truth areas are outlined in green. Note that BRACS does not provide detailed annotations, so ground truth is not displayed in panel (d). Additional visualizations can be found in the Appendix (Sec.~\ref{sec:app_qualitative}) for further reference. }
    \vspace{-5mm}
	\label{fig:qualitative}
\end{figure*}

We visualize and compare heatmaps generated by INSIGHT and baseline methods in Fig.~\ref{fig:qualitative}. Unlike baseline methods, which often rely on post-hoc adjustments, INSIGHT directly integrates interpretability into its architecture, mapping each patch’s probability to the heatmap. This eliminates the need for additional calibration steps, producing well-calibrated, fine-grained heatmaps that effectively highlight diagnostically significant regions. Additional visualizations are provided in Appendix~\ref{sec:app_qualitative}, which further illustrate the superior interpretability and precision of INSIGHT’s heatmaps across both WSIs and CT datasets.

\noindent \textbf{WSI Heatmap Comparisons:}
\begin{itemize}
    \item \textbf{INSIGHT:} The heatmaps generated by INSIGHT accurately capture subtle structures and ensure comprehensive coverage of ground-truth areas. Zoomed-in views reveal precise delineation of lesion boundaries, showcasing INSIGHT’s ability to localize regions critical for diagnosis. Ground-truth areas exhibit high activation values, reflecting the model’s robustness in spatially aligning predictions with annotations.
    \item \textbf{ABMIL:} Heatmaps from ABMIL frequently produce false negatives, detecting only a limited number of patches and failing to capture a holistic view of lesion regions. This lack of coverage diminishes its diagnostic utility, particularly in cases with subtle abnormalities.
    \item \textbf{CLAM:} While CLAM’s heatmaps overlap with ground-truth regions, they often exhibit poor calibration, with significant portions of the lesion areas showing low activation values. This leads to suboptimal localization, especially for boundaries.
    \item \textbf{TransMIL:} Heatmaps generated by TransMIL suffer from numerous false positives, marking non-lesion areas as relevant. This overactivation reduces diagnostic precision and makes the results less interpretable for clinical use.
    \item \textbf{WiKG:} Although WiKG captures interactions between patches through dynamic graph representation, its heatmaps occasionally exhibit overactivation in non-lesion areas. This impact boundary delineation and hurt interpretability in clinical applications.
\end{itemize}

\begin{figure*}[t]
	\centering
	\includegraphics[width=0.9\textwidth]{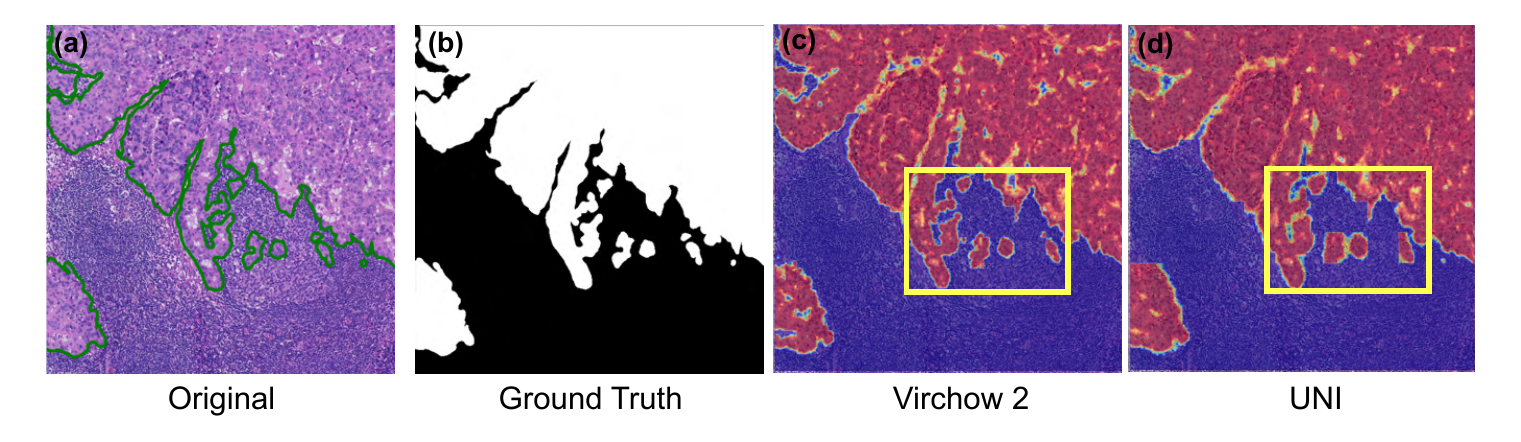}
    \vspace{-5mm}
	\caption{Qualitative comparison of heatmaps generated by INSIGHT using different foundation models. We compare lesion detection heatmaps generated from (a) the original slide, (b) the ground truth in binary (c) INSIGHT with UNI, and (d) INSIGHT with Virchow 2. Yellow box region highlights differences in boundary sharpness and focus.}
    \vspace{-5mm}
	\label{fig:virchow2_qual}
\end{figure*}

\noindent \textbf{Impact of Foundation Model.}
To further evaluate the generalizability of INSIGHT across different foundation models, we compare heatmaps generated using \textbf{Virchow 2} with those from the default \textbf{UNI} backbone. As shown in Fig.~\ref{fig:virchow2_qual}, both foundational models enable accurate localization of lesion regions; however, Virchow 2 produces slightly more refined boundaries in certain zoomed-in regions, suggesting that foundation model choice can affect visual precision. Despite this, INSIGHT’s overall interpretability and region alignment remain robust across backbones.

\subsection{Comparison with Grad-CAM}
\label{sec:gradcam-comparison}
To demonstrate the superiority of INSIGHT’s built-in heatmaps over post-hoc methods, we compare them with Grad-CAM on the MosMed and CAMELYON16 datasets (Table~\ref{tab:gradcam-comparison}). Unlike Grad-CAM, which requires additional backward passes and is sensitive to noisy gradients, INSIGHT generates interpretable heatmaps directly during forward inference without extra computation. The results show that INSIGHT achieves significantly higher Dice scores across all datasets and foundation models, indicating more accurate and fine-grained localization of diagnostically relevant regions. This intrinsic explainability ensures better alignment with clinical expectations while avoiding the instability and coarse resolution often observed in Grad-CAM.

\subsection{Ablation Studies}
\label{sec:ablation-studies}
To validate the effectiveness of INSIGHT’s architectural components, we perform an ablation study analyzing three key innovations: context suppression, SmoothMax pooling, and regularization. Results for the CAMELYON16 and BRACS datasets are shown in Table~\ref{table:combined_ablation}. The study evaluates both binary (CAMELYON16) and multi-label (BRACS) tasks, revealing how each component contributes to classification and segmentation performance. Our key findings are the following. 
\begin{table}[t]
\centering
\caption{Comparison of Dice scores (mean $\pm$ std) between INSIGHT’s built-in heatmaps and Grad-CAM. Results are reported on the MosMed and CAMELYON16 datasets.}
\renewcommand{\arraystretch}{1.2}
\setlength{\tabcolsep}{10pt}
\small
\resizebox{0.8\textwidth}{!}{
\begin{tabular}{l l | c c}   
\toprule
\textbf{Dataset} & \textbf{Foundation Model} & \textbf{Built-in Heatmap} & \textbf{Grad-CAM} \\
\midrule
MosMed     & DINOv2   & \textbf{42.7 $\pm$ 15.3} & 12.8 $\pm$ 13.9 \\
Camelyon16 & UNI      & \textbf{74.6 $\pm$ 19.1}          & 26.4 $\pm$ 20.4 \\
Camelyon16 & Virchow2 & \textbf{78.3 $\pm$ 15.4}          & 20.6 $\pm$ 27.4 \\
\bottomrule
\end{tabular}
}
\label{tab:gradcam-comparison}
\end{table}

\begin{table}[t]
\centering
\caption{Ablation study results on the CAMELYON16 and BRACS datasets. 
INSIGHT performance is shown with different configurations of CS (Context Suppression), 
SM (SmoothMax), and Rg (Regularizer).}
\renewcommand{\arraystretch}{1.2} 
\setlength{\tabcolsep}{10pt}    
\small
\resizebox{0.8\textwidth}{!}{
\begin{tabular}{ccc|cc|cccc}
\toprule
 &  &  & \multicolumn{2}{c|}{\textbf{CAMELYON16}} & \multicolumn{4}{c}{\textbf{BRACS}} \\
\cmidrule(lr){4-5}\cmidrule(lr){6-9}
\textbf{CS} & \textbf{SM} & \textbf{Rg} 
& \textbf{AUC} & \textbf{Dice(\%)} 
& \textbf{ADH} & \textbf{FEA} & \textbf{DCIS} & \textbf{Invasive} \\
\midrule
\xmark & \xmark & \xmark 
& \textbf{0.992} & 25.5$\pm$16.3 
& \textbf{0.741} & 0.509 & 0.764 & 0.993 \\
\cmark & \xmark & \xmark 
& 0.982 & 42.1$\pm$21.4 
& 0.733 & 0.507 & 0.799 & 0.998 \\
\cmark & \cmark & \xmark 
& 0.969 & \textbf{76.7$\pm$15.7} 
& 0.728 & 0.767 & 0.834 & 0.999 \\
\cmark & \cmark & \cmark 
& 0.990 & 74.6$\pm$19.1
& 0.734 & \textbf{0.790} & \textbf{0.837} & \textbf{0.999} \\
\bottomrule
\end{tabular}
}
\label{table:combined_ablation}
\end{table}

\noindent  \textbf{Context Suppression.} Adding context suppression (Row 2) improves Dice scores across all datasets, with a notable gain of $16.6$ on CAMELYON16. This highlights the module’s ability to improve spatial alignment between predictions and ground truth by focusing on relevant regions while suppressing background noise. However, context suppression alone does not maximize AUC, indicating spatial coherence is insufficient for capturing the discriminative features needed for optimal classification.

\noindent \textbf{SmoothMax Pooling.} Incorporating SmoothMax pooling alongside context suppression (Row 3) yields a substantial Dice increase of $34.6$ on CAMELYON16, emphasizing its role in refining lesion boundaries by leveraging all pixel-level activations. This pooling strategy enhances the model’s ability to capture nuanced spatial details and boundary precision. However, it slightly reduces AUC for binary tasks, likely due to softened decision boundaries introduced by spatial averaging, which may blur class distinctions. Its ablation (“SmoothMax Removal”) reverts to standard max pooling.

\noindent \textbf{Regularization.} Adding regularization (Row 4) improves both AUC and Dice across datasets, stabilizing AUC in CAMELYON16 and enhancing robustness in BRACS. Spectral decoupling mitigates overfitting in single-label tasks like CAMELYON16, ensuring broader feature utilization and improved generalization. For multi-label tasks like BRACS, label smoothing particularly benefits smaller classes such as ADH and FEA, improving classification consistency across imbalanced distributions.

\noindent \textbf{Ablation with Virchow2 Backbone. }
To further evaluate the generalizability of INSIGHT’s architectural components, we repeat our ablation study using the Virchow2~\citep{zimmermann2024virchow2scalingselfsupervisedmixed} foundational model. This experiment is conducted on CAMELYON16 to assess how our modules perform under a different pretraining regime.
As shown in Table~\ref{tab:virchow2_comparison}, we observe consistent performance gains with the addition of each module. The full INSIGHT configuration, incorporating context suppression, SmoothMax pooling, and regularization, achieves the highest overall performance with an AUC of $0.992$ and a Dice score of $78.3\%$. These results reinforce the robustness of our architecture and demonstrate that INSIGHT’s design principles are effective across diverse pretrained backbones.
\begin{table}[t]
    \centering
    \caption{Ablation study of INSIGHT using the Virchow2 foundational model on the Camelyon16 dataset.}    \renewcommand{\arraystretch}{1.2}
    \small
    \resizebox{0.4\textwidth}{!}{
    \begin{tabular}{ccc|cc}
        \toprule
        \multicolumn{3}{c|}{} & \multicolumn{2}{c}{\textbf{Virchow2}} \\
        \cmidrule{4-5}
        \textbf{CS} & \textbf{SM} & \textbf{Rg} & \textbf{AUC} & \textbf{Dice (\%)} \\
        \midrule
        \xmark & \xmark & \xmark & 0.968 & 3.8$\pm$3.0 \\
        \cmark & \xmark & \xmark & 0.988 & 34.8$\pm$21.2 \\
        \cmark & \cmark & \xmark & 0.990 & 78.0$\pm$17.3 \\
        \cmark & \cmark & \cmark & \textbf{0.992} & \textbf{78.3$\pm$15.4} \\
        \bottomrule
    \end{tabular}
    }
    \label{tab:virchow2_comparison}
\end{table}

\section{Discussion and Conclusion} 

Accurately detecting and interpreting disease-related regions in medical imaging remains a critical challenge, particularly when only image-level annotations are available. INSIGHT addresses this gap by achieving state-of-the-art performance in classification and segmentation across CT volumes and WSIs using \textit{only} image-level labels. This capability is transformative, aligning with real-world clinical workflows where physicians rarely annotate at the pixel or voxel level. By enabling use of larger datasets, INSIGHT unlocks potential for scalable, efficient medical imaging solutions. Its architectural innovations—detection and context modules—combine local feature analysis with contextual understanding to produce interpretable, well-calibrated heatmaps. These heatmaps enhance diagnostic transparency and reduce reliance on pixel-level annotations. Such capabilities could significantly accelerate the annotation process~\citep{raciti2020novel}, allowing clinicians to refine rather than annotate from scratch, streamlining workflows requiring detailed labels.

\noindent \textbf{Limitations.}
While INSIGHT has been validated on CT and WSI data, its applicability to other modalities, such as MRI and ultrasound, remains unexplored. In principle, the architecture should generalize due to its ability to process spatially embedded features and generate interpretable outputs. However, a key barrier lies in the scarcity of public datasets. Future work will focus on identifying or curating suitable datasets to assess generalizability across broader modalities.

\noindent \textbf{Conclusion.}
INSIGHT represents a significant advancement in weakly supervised medical imaging, demonstrating strong performance across diverse tasks and modalities. By leveraging image-level labels, embedding interpretability directly into its architecture, and providing fine-grained diagnostic insights, INSIGHT lays the foundation for scalable, clinically relevant AI solutions. Future work will enhance its capabilities with advanced foundation models and extend its reach to additional modalities, broadening its impact on medical imaging research and practice.

\acks{This work was supported in part by NSF award \#2326491. The views and conclusions contained herein are those of the authors and should not be interpreted as the official policies or endorsements of any sponsor. We thank Jhair Gallardo and Shikhar Srivastava for their comments on early drafts.}

\bibliography{library}

\newpage
\appendix

\begin{center}
    {\Large{\textbf{Appendix}}}
\end{center}

\section{Implementation Details}
\label{appendix:training}
\noindent \textbf{Architecture Parameters.} INSIGHT uses UNI~\citep{chen2024uni} and Virchow~\citep{vorontsov2024foundation} for WSI tiles, and ViT-L/DINOv2~\citep{oquab2024dinov2} for CT slices. Pre-processed inputs are resized to $224 \times 224$, and embeddings from the last layer of the encoder are concatenated to retain spatial information. Dimensionality is reduced from $1024$ to $128$ using a linear transformation. Detection and Context modules consist of three convolutional layers with GELU activations, layer normalization, and a final layer for label-specific outputs.
The Detection module uses $1 \times 1$ convolutions, while the Context module uses $3 \times 3$ convolutions. 

\noindent \textbf{Training Configuration.} INSIGHT is trained using AdamW with a learning rate of $10^{-4}$ and weight decay of $10^{-4}$. Training is capped at $50$ epochs, with early stopping triggered after $8$ epochs of no validation improvement. We performed grid search to tune hyperparameters on the Camelyon16 validation set, including SmoothMax sharpness $\alpha$ (4, 6, \textbf{8}, 10), learning rates (5e$^{-5}$, \textbf{1e$^{-4}$}, 3e$^{-4}$), and spectral decoupling strengths $\lambda$ (0, \textbf{0.01}, 0.05). The selected configuration was then applied to BRACS and MosMed without further tuning, demonstrating INSIGHT’s robustness across tasks. To ensure fair comparison, all baseline models were implemented using the original configurations specified in their respective papers.

\section{Qualitative Analysis}
\label{sec:app_qualitative}
\begin{figure*}[h!]
    \centering
    \includegraphics[width=0.99\textwidth]{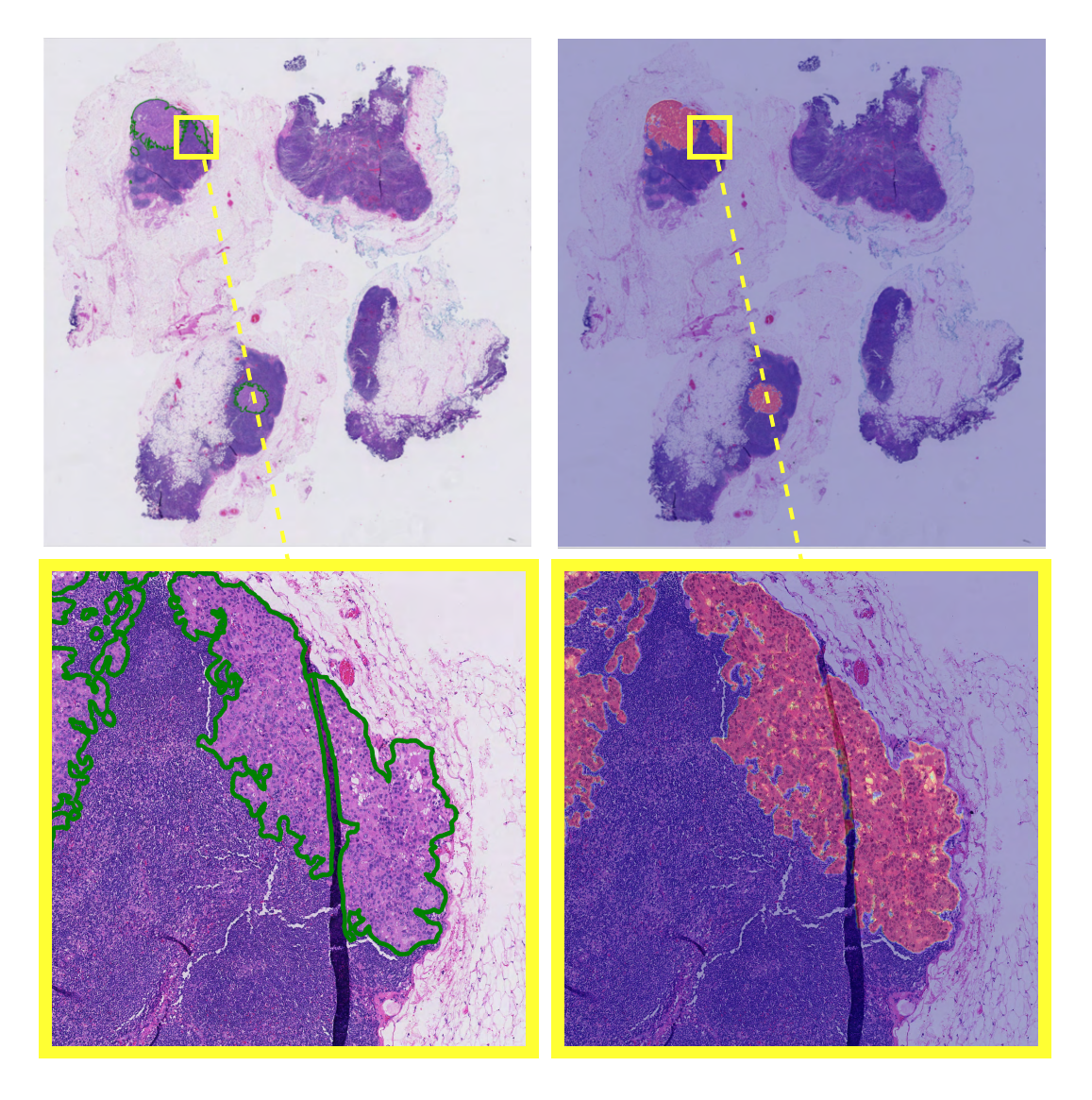}
     \caption{Original histopathology image (left) and the corresponding heatmap generated by INSIGHT using the \textbf{Virchow 2} encoder (right). The green outlines on the original image represent ground-truth regions. The bottom row provides zoomed-in views of the regions highlighted by yellow boxes.}   
    \label{fig:supplementray_results_1}
\end{figure*}

\begin{figure*}[t]
    \centering
    \includegraphics[width=0.99\textwidth]{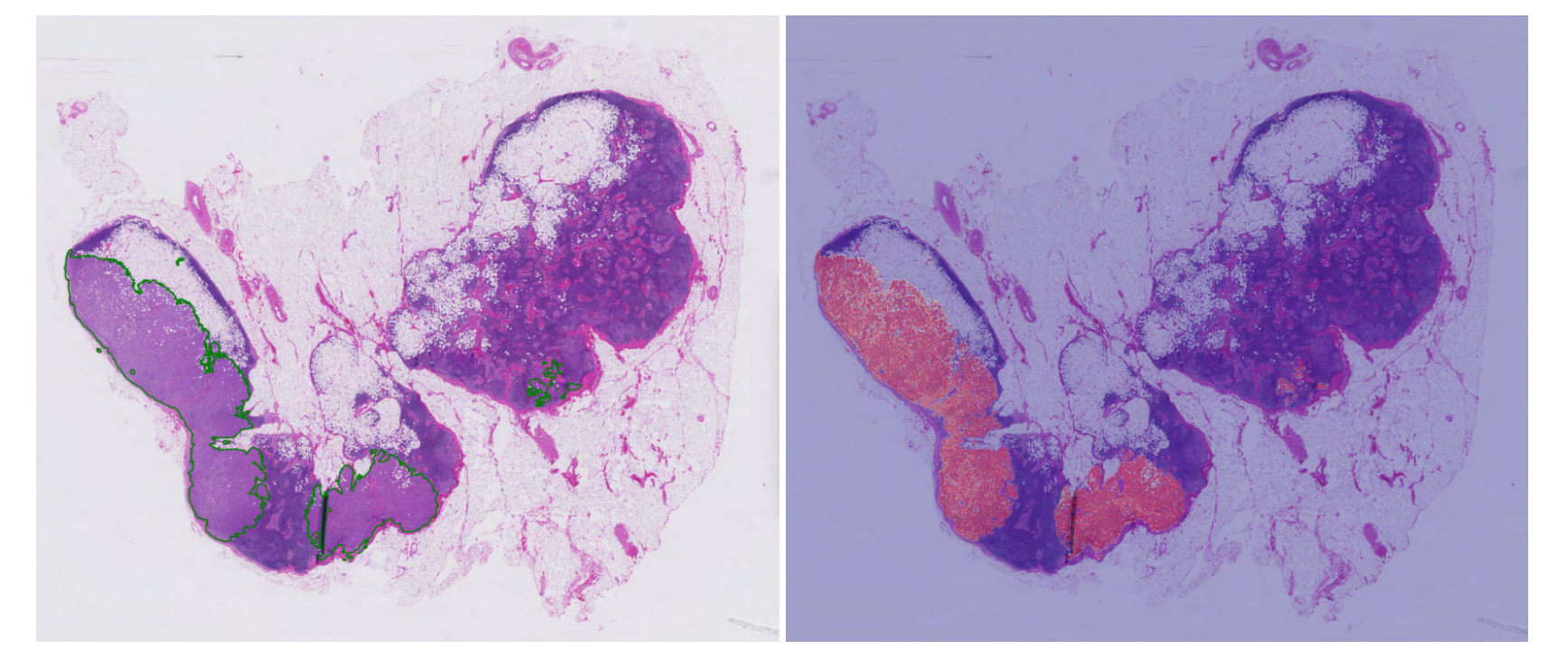}
    \caption{WSI heatmap visualization (right) generated by INSIGHT using \textbf{Virchow 2} as the feature extractor on the CAMELYON16 official dataset. The green outlines on the original image (left) represent ground-truth regions.}
    \label{fig:supplementray_results_2}
\end{figure*}

\begin{figure*}[t]
    \centering
    \includegraphics[width=0.99\textwidth]{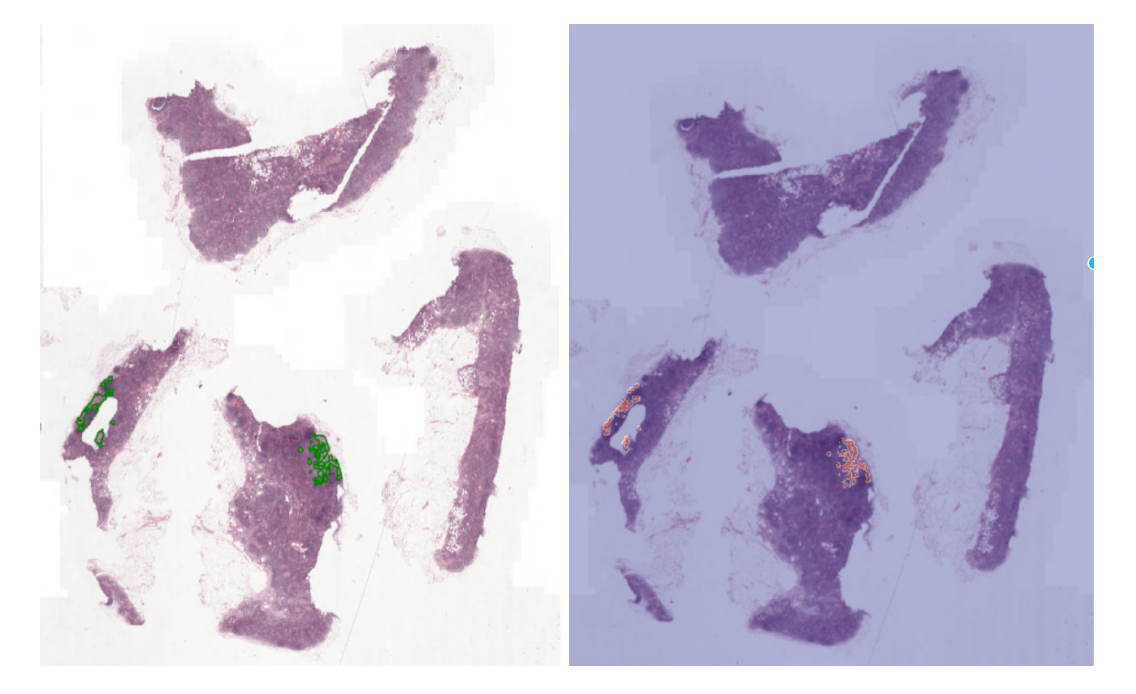}
    \caption{WSI heatmap visualization (right) generated by INSIGHT using \textbf{UNI} as the feature extractor on the CAMELYON16 official dataset. The green outlines on the original image (left) represent ground-truth regions.}
    \label{fig:supplementray_results_3}
\end{figure*}

\begin{figure*}[h!]
    \centering
    \includegraphics[width=0.99\textwidth]{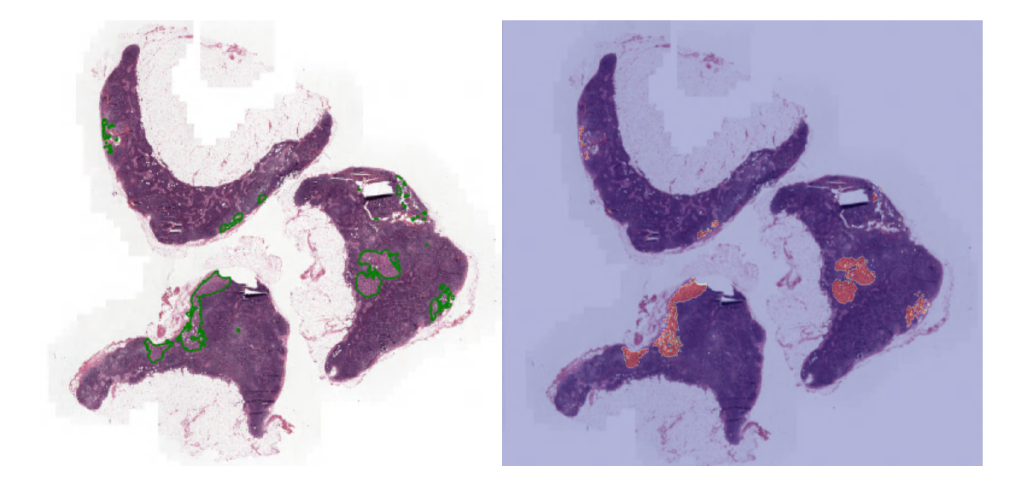}
    \caption{Additional WSI heatmap visualization (right) generated by INSIGHT using \textbf{UNI} as the feature extractor on the CAMELYON16 official dataset. The green outlines on the original image (left) represent ground-truth regions.}
    \label{fig:supplementray_results_4}
\end{figure*}

We provide additional visualization results (Fig.~\ref{fig:supplementray_results_1},~\ref{fig:supplementray_results_2},~\ref{fig:supplementray_results_3},~\ref{fig:supplementray_results_4}) to further demonstrate INSIGHT’s ability to generate well-calibrated heatmaps that align with diagnostically relevant regions in the CAMELYON16 dataset. These examples include results using both UNI and Virchow 2 feature encoders, highlighting the consistency and generalizability of our method across different backbone models. Please zoom in to view finer details.

\clearpage

\section{Runtime Analysis}
\label{sec:app_runtime}

A key advantage of INSIGHT is native support for multi-label classification. In contrast, baselines like ABMIL and CLAM variants are inherently multi-class only and require running separate models per label for multi-label tasks such as BRACS. This makes INSIGHT significantly more efficient in real-world settings. We report runtime (min/epoch) for training and inference below.

\begin{table}[ht]
\centering
\caption{Runtime (min/epoch) comparison across datasets and models.}
\label{tab:runtime}
\resizebox{0.6\textwidth}{!}{%
\begin{tabular}{l l c c}
\toprule
\textbf{Dataset (task)} & \textbf{Model} & \textbf{Train} & \textbf{Inference} \\
\midrule
\multirow{6}{*}{Camelyon16 (binary)} 
 & INSIGHT (Ours)    & 81.8  & 4.2 \\
 & ABMIL      & 21.2           & 2.2 \\
 & CLAM-SB    & 19.3           & 2.2 \\
 & CLAM-MB    & 21.5           & 2.3 \\
 & TransMIL   & 28.3           & 2.6 \\
 & WiKG       & 24.7           & 2.4 \\
\midrule
\multirow{6}{*}{BRACS (multilabel)} 
 & INSIGHT (Ours)     & 114.7 & 6.8 \\
 & ABMIL      & 111.5          & 12.8 \\
 & CLAM-SB    & 112.8          & 15.3 \\
 & CLAM-MB    & 121.4          & 16.6 \\
 & TransMIL   & 132.7          & 19.2 \\
 & WiKG       & 125.3          & 16.9 \\
\bottomrule
\end{tabular}
}
\end{table}

\clearpage
\section{Heatmap Generation Algorithm}
\label{appendix:algorithm}

\begin{algorithm}[h]
\caption{INSIGHT Heatmap Generation and Classification}
\label{alg:heatmap}
\SetAlgoLined
\DontPrintSemicolon
\KwIn{WSI tiles or CT slices $\{x_i\}_{i=1}^N$ with embeddings $\{e_i\}_{i=1}^N$, Detection module $f_{\text{Det}}$, Context module $f_{\text{Con}}$, and pooling sharpness $\alpha$.}
\KwOut{Prediction $\hat{y}$ and reconstructed heatmap $\mathcal{H}_\text{full}$.}

\BlankLine
\textbf{Step 1: Feature Projection.} \\
\For{$i = 1$ \KwTo $N$}{
    $e_i \gets \text{Conv}_{1 \times 1}(e_i)$ \tcp*{1×1 projection to reduce dimension}
}

\BlankLine
\textbf{Step 2: Module Outputs.} \\
\For{$i = 1$ \KwTo $N$}{
    $\mathcal{H}_{\text{Det},i} \gets f_{\text{Det}}(e_i)$ \tcp*{Local detail map}
    $\mathcal{H}_{\text{Con},i} \gets f_{\text{Con}}(e_i)$ \tcp*{Context map}
    $\mathcal{H}_i \gets \sigma \big( (1 - \sigma(\mathcal{H}_{\text{Con},i})) \odot \mathcal{H}_{\text{Det},i} \big)$
}

\BlankLine
\textbf{Step 3: Patch-to-Image Fusion.} \\
Arrange $\{\mathcal{H}_i\}$ into a full heatmap $\mathcal{H}_{\text{full}}$ according to spatial coordinates. \\
Apply Otsu threshold $T \gets \arg \min_t \sigma_w^2(t)$. \\
$\mathcal{H}_{\text{full}} \gets \mathcal{H}_{\text{full}} \cdot \mathbb{I}(\mathcal{H}_{\text{full}} > T)$

\BlankLine
\textbf{Step 4: SmoothMax Pooling.} \\
$\hat{y} \gets \frac{\sum_{j} \mathcal{H}_{\text{full}, j} \cdot \exp(\alpha \cdot \mathcal{H}_{\text{full}, j})}{\sum_{j} \exp(\alpha \cdot \mathcal{H}_{\text{full}, j})}$

\BlankLine
\textbf{Step 5: (Multi-label).} \\
If multi-label, repeat Steps 2–4 for each class $c \in \{1, \dots, C\}$, where $C$ is the number of labels. This produces channel-wise outputs $\hat{y}_c$, resulting in a prediction vector $\hat{\mathbf{y}} \in \mathbb{R}^C$.

\end{algorithm}

\end{document}